\documentclass[%
reprint,
superscriptaddress,
groupedaddress,
nobibnotes,
amsmath,amssymb,
aps,
prl,
]{revtex4-2}

\usepackage[dvipsnames]{xcolor}
\usepackage{graphicx}
\usepackage{dcolumn}
\usepackage{bm}
\usepackage[breaklinks=true,colorlinks,citecolor=blue,linkcolor=red,urlcolor=blue]{hyperref}
\usepackage{braket}
\usepackage[utf8]{inputenc}
\usepackage{mathtools}
\usepackage[bottom]{footmisc}
\usepackage{amsfonts}
\usepackage{slashed}
\usepackage{amsthm}
\usepackage{tikz}
\usepackage{booktabs}

\usepackage{pifont}
\usepackage{graphicx}

\begin{document}

\title{Breakdown of Topological Inheritance and Twist-Induced Quantum Geometry Reconfiguration in Moiré Flat Bands}
\author{Yi-Chun Hung}

\author{Xiaoting Zhou}
\email[Contact author: ]{x.zhou@northeastern.edu}

\author{Arun Bansil}
\email[Contact author: ]{ar.bansil@northeastern.edu}

\affiliation{Department of Physics,\;Northeastern\;University,\;Boston,\;Massachusetts\;02115,\;USA}
\affiliation{Quantum Materials and Sensing Institute,\;Northeastern University,\;Burlington,\;Massachusetts\;01803,\;USA}


\begin{abstract}
Since the inception of moiré quantum matter, a foundational tenet of the field has been that the quantum geometry of emergent flat bands is faithfully inherited from the low-energy valleys of the constituent monolayers. Here, we demonstrate a breakdown of this longstanding tenet in twisted bilayers of loop-current-ordered kagome lattices (tb-LCK). Using microscopic tight-binding modeling, we reveal a twist-induced reconfiguration of quantum geometry where realistic interlayer hybridization quenches topological inheritance from the monolayer. By tuning the loop-current phase, we identify distinct regimes in which the monolayer Berry curvature is either substantially redistributed or entirely suppressed in the moiré flat bands. We further show that this quantum geometric collapse is expected to be readily accessible in vanadium-based kagome metals such as AV$_{\text{3}}$Sb$_{\text{5}}$, and that Floquet engineering via waveguide laser illumination offers a practical route to turn topological inheritance on and off. Our findings uncover a universal mechanism for quantum geometric reconstruction, establishing interlayer coupling strength as an independent parameter for tuning band topology beyond the weakly coupled van der Waals heterostructure paradigm.
\end{abstract}

\maketitle

Twisted van der Waals heterostructures have revolutionized the control of strongly correlated and topologically nontrivial phases. The long-wavelength superlattice potential arising from the superposition of crystalline lattices profoundly reshapes the electronic dispersion, generating flat bands characterized by strongly enhanced interactions and a rich landscape of quantum geometry. These features underpin the exotic physics observed in twisted bilayer graphene (TBG) \cite{Andrei2020} and transition-metal dichalcogenides (tb-TMDs) \cite{Devakul2021}, yielding discoveries ranging from unconventional superconductivity \cite{Cao2018, Balents2020} to fractional quantum anomalous Hall states \cite{Cai2023, Park2023, PhysRevX.13.031037}.

Central to the field has been the principle of topological inheritance, which holds that the moiré potential acts as a gentle perturbation that folds the monolayer bands into a mini-Brillouin zone (mBZ). In TBG as well as tb-TMDs, the quantum geometric properties of the moiré flat bands, such as concentrated Berry curvature and nontrivial Wilson-loop windings, are directly inherited from the low-energy states of the unhybridized monolayers \cite{PhysRevLett.124.167002, PhysRevLett.128.087002, PhysRevLett.122.106405, PhysRevB.99.155415}, providing a robust, predictive framework for engineering moiré bands. However, a key open question is whether this inheritance survives in more complex systems involving multi-orbital frustration or symmetry-breaking orders, where the energy scale of the interlayer coupling can compete with that of intrinsic band gaps. 

We address this question by investigating the twisted bilayer of a loop-current-ordered kagome lattice (tb-LCK). Vanadium-based kagome metals, AV$_3$Sb$_5$ (A = K, Rb, Cs), host intertwined charge-density waves (CDWs) and superconducting orders accompanied by spontaneous time-reversal symmetry (TRS) breaking \cite{Wilson2024, Mielke2022}. Theoretical models predict a $2\times2$ CDW order featuring a Star-of-David pattern alongside a complex loop-current (LC) order \cite{PhysRevB.106.144504, Tazai2023, PhysRevB.107.045127, Zhou2022}, analogous to the Haldane model \cite{PhysRevLett.61.2015}. In contrast to monolayer TMDs \cite{PhysRevLett.135.196402, calugaru2024MTMD, jiang20242database}, where the low-energy physics is governed by gapped Dirac cones at the $K/K'$ valleys \cite{Manzeli2017}, the monolayer-LCK band structure exhibits topological gaps with a nonzero Chern number and Berry curvature concentrated at $M$ valleys. This intrinsic topological obstruction invalidates conventional long-wavelength continuum models, demanding a full tight-binding (TB) framework.

\par In this Letter, we demonstrate that twisted LCK bilayers exhibit a striking \textit{reconfiguration of quantum geometry} that violates the principle of topological inheritance,. Using a TB framework, we show that for realistic interlayer tunneling ($t_z$) values, the highly localized Berry curvature of the parent layers is effectively quenched via hybridization with energetically distant states. This reconfiguration is driven by the competition between the energy scales of $t_z$ and the intrinsic topological band gaps. This competition gives rise to both a weak-coupling regime in which the inheritance is preserved and a distinct strong-coupling regime of geometric collapse. In this way, we identify tb-LCK as not merely a variant of TBG but as a new platform for topological band engineering, where the twist serves the control parameter for manipulating emergent topology and enabling quantum geometric phenomena beyond the conventional moiré paradigm.


\begin{figure}[t]
  \centering
  \centering
    \includegraphics[width=\linewidth]{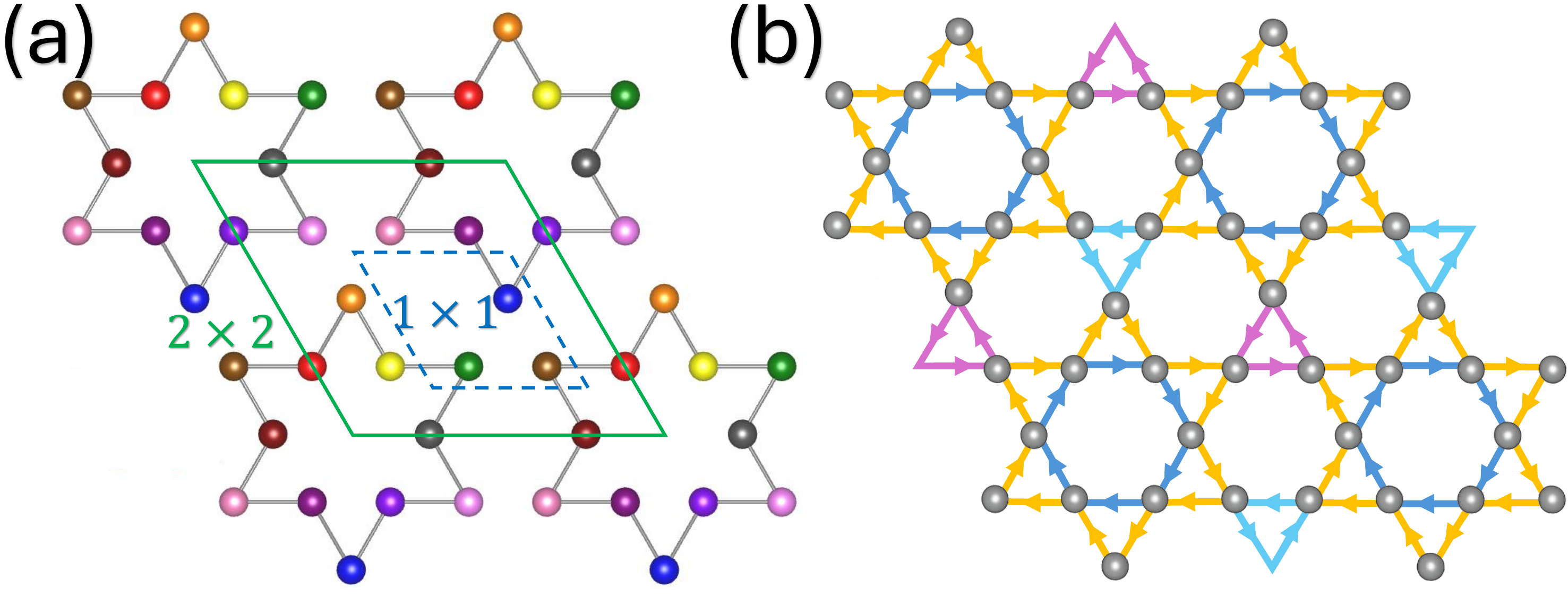}
  \caption{(a) Lattice structure of LCK lattice highlighting the Star-of-David pattern. Green solid and blue dashed lines mark the $2\times2$ and $1\times1$ cells, respectively. (b) A schematic of the loop currents in LCK. Distinct loops are colored differently.
  }
  \label{fig:01}
\end{figure}

\begin{figure}[t]
  \centering
  \centering
    \includegraphics[width=\linewidth]{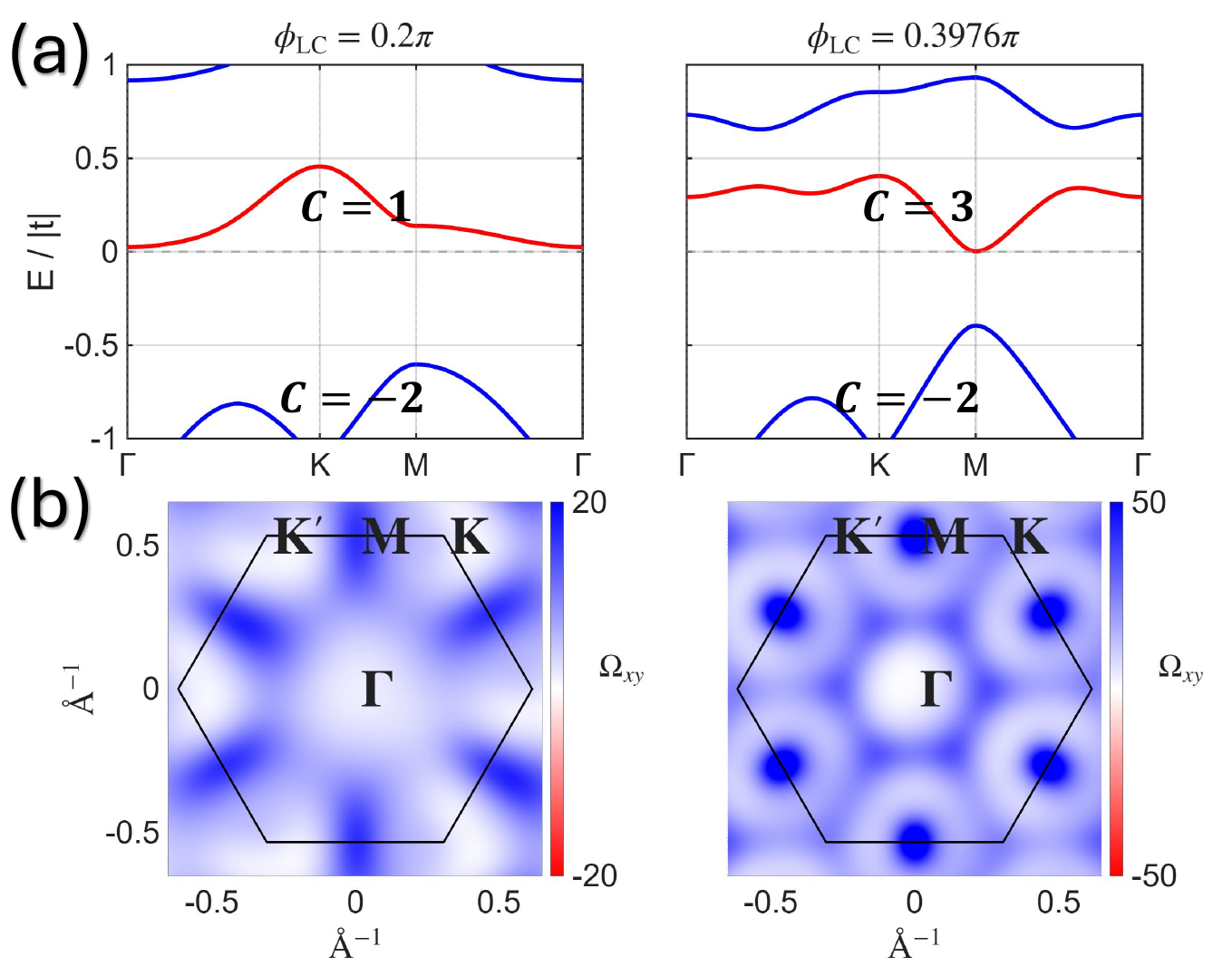}
  \caption{ (a) Band structures of LCK at $\phi_{\text{LC}} = 0.2\pi$ and $0.3976\pi$, with the 6th lowest band (B6) highlighted in red line and nearby bands labeled by their Chern numbers $C$. (b) The Berry curvature of B6.
  }
  \label{fig:02}
\end{figure}

\paragraph{Monolayer Loop-Current Kagome Lattice---}
We define the mean-field TB Hamiltonian for the $2 \times 2$ CDW order in LCK, which preserves $C_{6z}$ symmetry while breaking time-reversal symmetry (TRS) via its imaginary component. This generates a Star-of-David pattern (Fig.~\ref{fig:01}(a)) and directed loop currents (Fig.~\ref{fig:01}(b)):
\begin{align}
    & H_0 = -t\sum_{\langle\alpha\beta\rangle}\hat{C}_\alpha^\dagger\hat{C}_\beta - \mu\sum_{\alpha} \hat{C}_\alpha^\dagger\hat{C}_\alpha\notag \\ 
    & +\sum_{\mathbf{R}(\alpha\beta\gamma)}h(\mathbf{Q}_\gamma)\big(\hat{C}_{\mathbf{R},\alpha}^\dagger\hat{C}_{\mathbf{R},\beta} - \hat{C}_{\mathbf{R},\alpha}^\dagger\hat{C}_{\mathbf{R}-\mathbf{a}_\gamma,\beta} \big) + \text{H.c.},
\end{align}
where $h(\mathbf{Q}_\gamma)=t_{\text{CDW}}\cos(\mathbf{Q}_\gamma\cdot\mathbf{R})$. The lattice vector $\mathbf{R}=m\mathbf{a}_1+n\mathbf{a}_2$ specifies the position of the cell, with $m,n\in\mathbb{N}$ and $\mathbf{a}_i$ the $i$th primitive translation vector of the $1\times1$ unit cell. The CDW nesting vector is given by $\mathbf{Q}_\gamma=\mathbf{G}_\gamma/2$, where $\mathbf{G}_\gamma$ denotes the $\gamma$th reciprocal lattice vector.  $(\alpha,\beta,\gamma)\in\{1,2,3\}$ label the sublattices. $\langle \cdots \rangle$ denotes the nearest-neighbor (NN) pairs, and $(\alpha\beta\gamma)$ runs over cyclic permutations of $(123)$ \cite{Zhou2022}. We set the chemical potential at $\mu=0$ and the CDW strength $|t_{\text{CDW}}|\cong0.3162|t|$. Therefore, low-energy physics is governed by the 6th lowest band (B6 in Fig. 4), which crosses the Fermi level for various phases of the LC ordered phase $\phi_{\text{LC}}=\text{Arg}(t_{\text{CDW}})$. As Fig.~\ref{fig:02} shows, the location of the Berry curvature hotspots of band B6 varies sensitively with $\phi_{\text{LC}}$, setting the stage for moiré hybridization. To characterize quantum geometric properties of LCK, we evaluate the lower bound of quantum weight $\tilde{K}$ and the reduced Hall conductivity $\bar{\sigma}$, defined as:
\begin{align}
\tilde{K} & \equiv 2\pi\int d[\mathbf{k}]|\Omega_{xy}|, 
\\ \bar{\sigma} & \equiv 2\pi\int d[\mathbf{k}]\Omega_{xy}.
\end{align}
The former provides a lower bound to the quantum metric, and the latter equals the Chern number when quantized and captures topological phase transitions. Here, $d[\mathbf{k}]=d^2\mathbf{k}/(2\pi)^2$ is the integral measure and $\Omega_{xy}$ is the Berry curvature of the band of interest; see Supplemental Materials (SM) for details \cite{SM}\nocite{PhysRevX.14.011052, PhysRevLett.133.206602, PhysRevB.90.165139, PhysRevB.95.024515, Peotta2015, yu2024quantum, Phys.Rev.B.112.035410, Phys.Rev.B.112.235134}.

\begin{figure}[h]
  \centering
  \centering
    \includegraphics[width=0.7\linewidth]{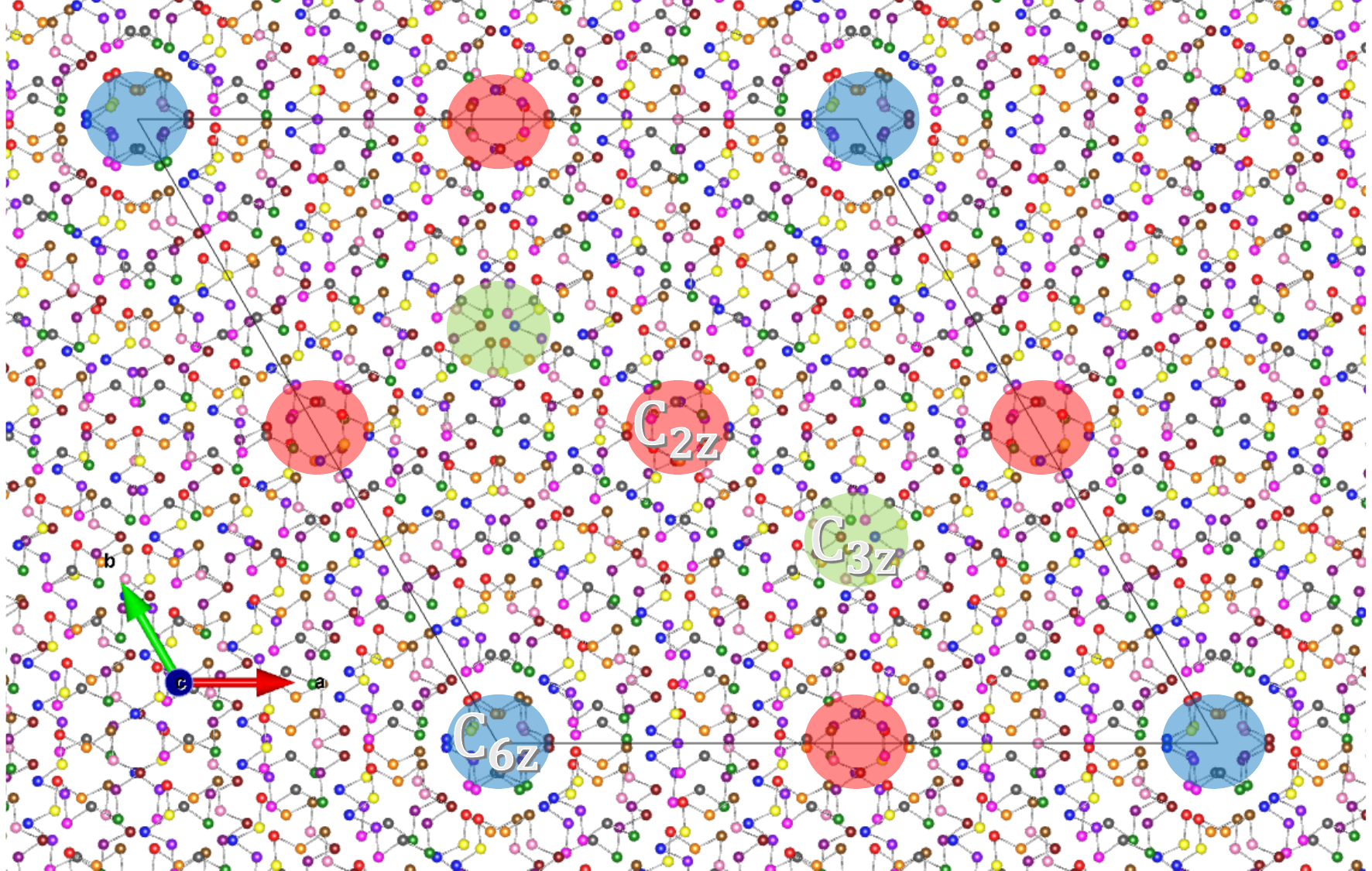}
  \caption{ Lattice structure of tb-LCK at $\theta_c \approx 9.43^\circ$, with high-symmetry stacking centers marked in different colors. See Fig.~S2 for structural details of various regions \cite{SM}.
  }
  \label{fig:03}
\end{figure}

\paragraph{Twist-Induced Reconfiguration in tb-LCK---}
The tb-LCK is constructed by rotating two LCK layers about the Star-of-David center, preserving the $C_{6z}$ symmetry. We model tb-LCK at commensurate twist angles $\theta_c$ using the Hamiltonian:
\begin{equation}
    H = H_{0,\frac{\theta_c}{2}}^{(t)} + H_{0,-\frac{\theta_c}{2}}^{(b)} + t_z\sum_{\substack{\alpha\in t \\ \beta\in b}}e^{-\lambda(\frac{r_{\alpha\beta}}{h}-1)}\hat{C}^\dagger_\alpha\hat{C}_\beta + \text{H.c.}.
\end{equation}
Here, $H_{0,\theta_c}^{(t,b)}$ denotes the monolayer TB Hamiltonian for the top ($t$) and bottom ($b$) layers, respectively.  $r_{\alpha\beta}=\|\mathbf{r}_\alpha-\mathbf{r}_\beta\|$ denotes the distance between the sublattice sites, while $h$ is the interlayer separation. The interlayer tunneling strength is set at $t_z=0.3|t|$, consistent with the reported values for layered kagome metals \cite{Ye2018, PhysRevB.111.045155}. A dimensionless decay length $\lambda=20$ is used to capture short-range van der Waals coupling \cite{PhysRevB.111.085137, PhysRevB.100.155421}. Commensurate twist angles are determined using $\theta_c=\cos^{-1}\!\left[(3m^2+3m+1/2)/(3m^2+3m+1)\right]$ for $m\in\mathbb{N}$ \cite{PhysRevLett.99.256802, PhysRevB.111.085137, PhysRevB.100.155421} without loss of generality. We have verified that lowering the threshold to $E_{\text{cut}}^{\text{(int)}} = 10^{-6}|t|$ yields negligible changes to the band structure. A schematic of the moiré superlattice is shown in Fig.~\ref{fig:03}. 

\par
\begin{figure}[t]
  \centering
  \centering
    \includegraphics[width=\linewidth]{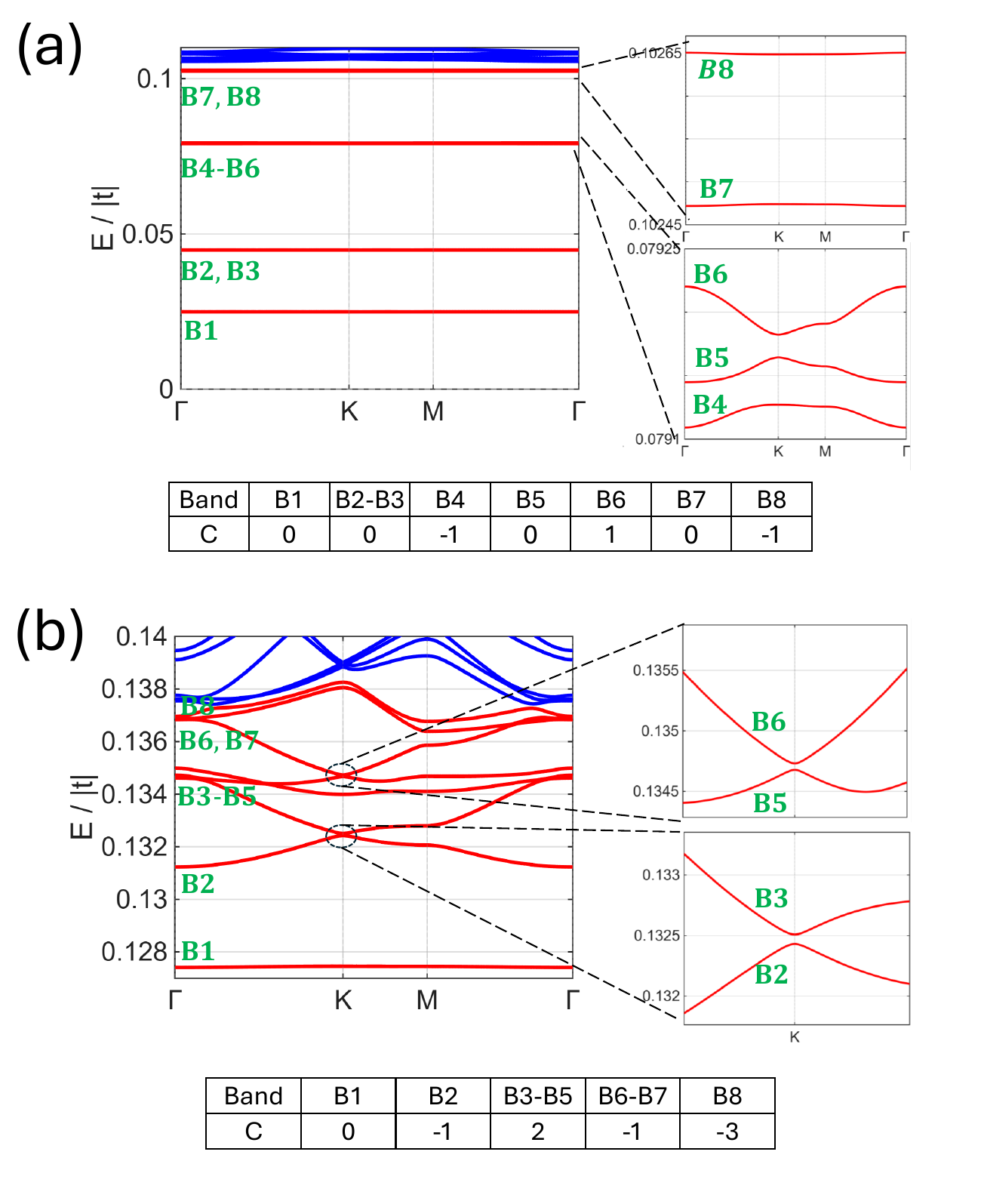}
  \caption{ Band structures of tb-LCK for $\phi_{\text{LC}} = 0.2\pi$ and (a) $t_z=0.3|t|$ and (b) $t_z=0.03|t|$ at $\theta_c\approx3.89^\circ$, where the lowest eight bands are marked in red lines with the indicated band indices. Insets highlight band gaps between the nearly degenerate bands. Chern numbers of various sets of bands are given in the tables.
  }
  \label{fig:04}
\end{figure}

\begin{figure}[t]
  \centering
  \centering
    \includegraphics[width=\linewidth]{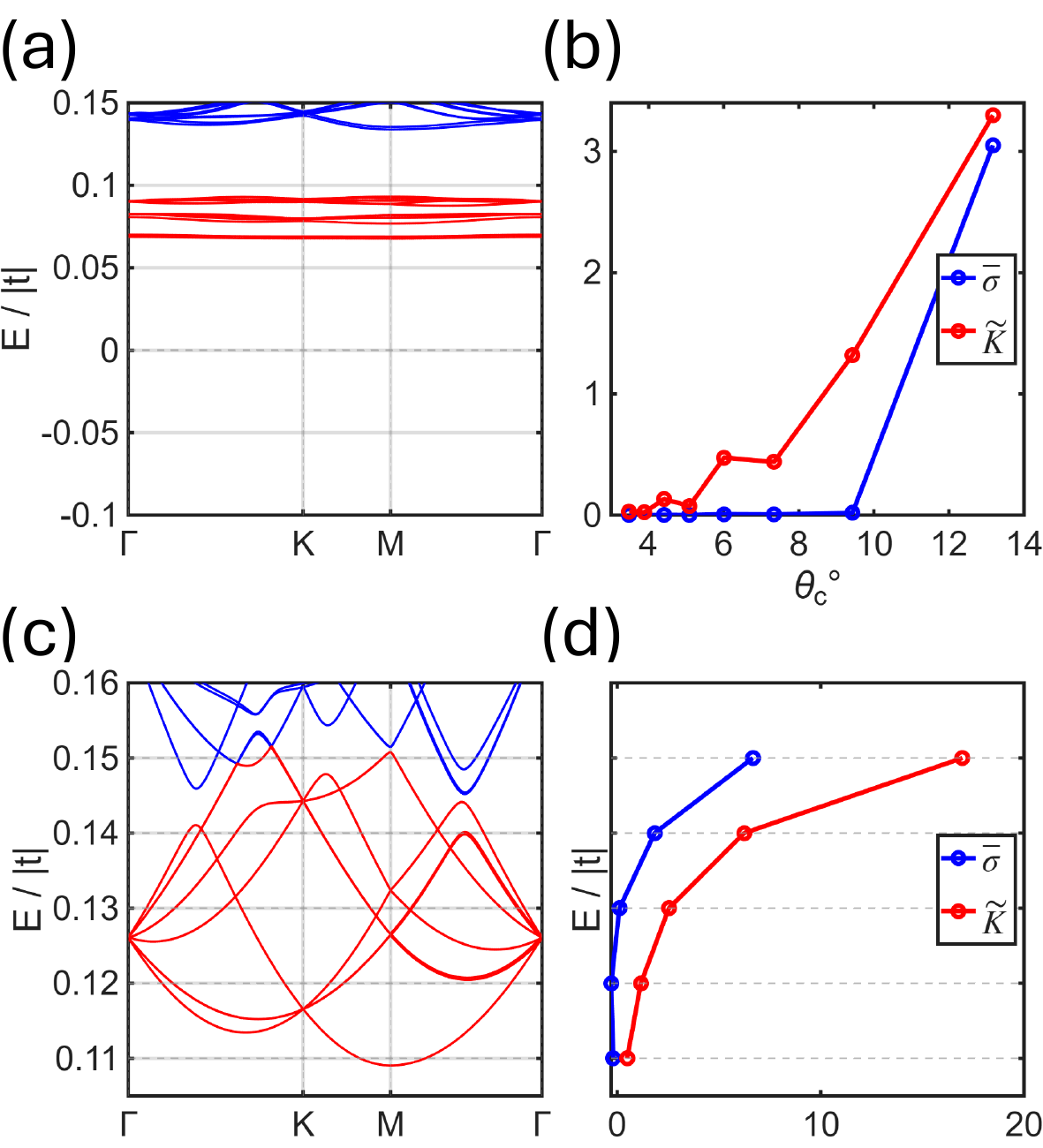}
  \caption{ Band structures of tb-LCK for $\phi_{\text{LC}} = 0.3976\pi$ and (a) $t_z=0.3|t|$ and (c) $t_z=0.03|t|$ at $\theta_c\approx3.89^\circ$, where the bands of interest are highlighted in red lines and labeled by their lower bound of quantum weight $\tilde{K}$. (b) $\bar{\sigma}$ and $\tilde{K}$ for all (twelve) low-energy bands of tb-LCK with $t_z=0.3|t|$ as functions of $\theta_c$. (d) $\bar{\sigma}$ and $\tilde{K}$ of tb-LCK with $t_z=0.03|t|$ at $\theta_c\approx3.89^\circ$ computed at selected chemical potentials marked by gray dashed lines in (c) at $k_BT\approx8.6\times10^{-3}|t|$.
  }
  \label{fig:05}
\end{figure}

 We turn now to discuss the evolution of moiré quantum geometry as a function of $\phi_{\text{LC}}$ and show a stark decoupling between the topological origin of the monolayer and the emergent moiré geometry. In the conventional moiré paradigm, shifts in the monolayer Berry curvature drive the topology of the twisted system. Here, however, the low energy moiré states frequently lack the quantum geometry of the monolayer, effectively suppressing topological inheritance. This nontrivial reorganization, driven by strong interlayer tunneling $t_z$, signifies a fundamental breakdown of the standard moiré paradigms. Representative results for the lower edge of band B6 at $\theta_c \approx 3.89^\circ$ (Figs.~\ref{fig:04} and \ref{fig:05}) confirm that this decoupling persists across a broad parameter space, most notably near $\phi_{\text{LC}} \approx 0.2\pi$ and $0.3976\pi$.

\paragraph{Hybridization-Driven Topological Suppression---}
This decoupling mechanism is vividly illustrated at $\phi_{\text{LC}} \approx 0.2\pi$. In the monolayer, the B6 band forms a $C = 1$ Chern band driven by a $\Gamma$-point band inversion, featuring a band edge at $\Gamma$, yet its Berry curvature is highly concentrated at the $M$ point (Fig.~\ref{fig:02}). This momentum-spatial separation between the topological origin and the Berry curvature hotspots indicates that the inherited quantum geometry cannot be captured by simple band-edge descriptions, leaving it being very sensitive to twist-induced interlayer scattering. Accordingly, upon twisting ($\theta_c \approx 3.89^\circ$), we observe that in the realistic coupling regime ($t_z=0.3|t|$) topological features are completely washed out, rendering the vast majority of the emergent low-energy moiré flat bands quantum-geometrically trivial (Fig.~\ref{fig:04}(a)). The mechanism underlying this collapse involves energy-scale competition: strong interlayer tunneling drives hybridization with energetically distant states, thoroughly disrupting the $\Gamma$-derived band edge character. Notably, when we artificially decrease the coupling to the weakly interacting regime ($t_z=0.03|t|$), the topological inheritance is restored, and an abundance of nontrivial moiré flat bands reappears (Figs.~\ref{fig:04}(b), S6, and S11 \cite{SM}).

\paragraph{Valley Quantum Geometry Collapse---}
We now extend our analysis to $\phi_{\text{LC}} \approx 0.3976\pi$, a regime predicted to host exotic correlated phases \cite{wang2025rotonSC}. As $\phi_{\text{LC}}$ increases, cascading band inversions endow the monolayer B6 band with a quadratic dispersion and sharply localized Berry curvature at the $M$ valleys (Figs.~\ref{fig:02} and S4 \cite{SM}). Despite this enhanced quantum geometry in the monolayer, we find that twisting drives a rapid topological collapse in the moiré bands at small twist angles.

\par By tracking the electronic structure across commensurate twist angles from $\theta_c \approx 13.17^\circ$ down to $3.48^\circ$, we can identify a 12-band low energy manifold originating from the lower band edge of band B6 (Figs.~\ref{fig:05}(a) and S7 \cite{SM}). A nontrivial topological phase ($C = -3$) survives only at the largest twist angle ($\theta_c \approx 13.17^\circ$), where the Berry curvature smears across generic $k$-points (Fig.~S8 \cite{SM}). As $\theta_c$ decreases, these bands become trivial, accompanied by a rapid decay of the lower bound of quantum weight $\tilde{K}$, reflecting a total loss of quantum geometry (Fig.~\ref{fig:05}(b)). This collapse is also driven by energy-scale competition: the realistic interlayer tunneling ($t_z=0.3|t|$) is comparable to the monolayer band gap separating bands B6 and B5, which have opposite signs of their Chern numbers (see Fig.~\ref{fig:02}(b)). The moiré potential thus violently hybridizes these opposite-Chern sectors, suppressing the low-energy quantum geometry.  

\par It is interesting to consider effects of artificially reducing the strength of interlayer tunneling to $t_z=0.03|t|$. This weak-coupling regime restores significant quantum geometry ($\bar{\sigma}$ and $\tilde{K}$) to the 12-band low-energy manifold at $\theta_c \approx 3.89^\circ$, despite strict energetic isolation from the higher energy bands (Figs.~\ref{fig:05}(c) and (d)). This recovery proves that the quantum-geometry collapse is dictated by the magnitude of interlayer hybridization, rather than the intrinsic moiré superlattice periodicity resulting from the twist alone. To establish experimentally accessible signatures of the CDW phase, we compute the finite-temperature responses of $\bar{\sigma}$ and $\tilde{K}$ by weighting the Berry curvature with the Fermi-Dirac distribution, see SM \cite{SM}. We evaluate these quantities at the $k_BT \approx 8.6\times10^{-3}|t|$ energy scale scale corresponding to transition temperature $T_c \sim 100K$, which is representative of hopping amplitudes ($t \sim 0.1$--$1$\,eV) in vanadium-based kagome metals \cite{PhysRevB.111.235114}. These parameters ensure that the calculated responses accurately reflect the quantum geometry near the phase transition. We emphasize that even though thermal redistribution of the Berry curvature can induce a sign change in $\bar{\sigma}$ at ultralow temperatures ($k_BT \approx 8.6\times10^{-7}|t|$), the restoration of the quantum geometry (reflected in the values $\tilde{K}$ and $\bar{\sigma}$) indicates that it is highly robust against thermal smearing \cite{SM}.

\paragraph{Experimental Realization---}
The tb-LCK physics unveiled in this study finds a promising experimental platform in the vanadium-based kagome metals AV$_3$Sb$_5$ (A = K, Rb, Cs), where the underlying kagome lattices are coupled via interlayer van der Waals interaction. These materials exhibit a $2 \times 2$ CDW order below $T_c\sim100$K \cite{PhysRevX.11.031026, PhysRevMaterials.5.034801, PhysRevLett.125.247002, Yin_2021, Luo2022, Jiang2021, Chen2021}, accompanied by spontaneous TRS breaking \cite{Mielke2022, yu2021hiddenflux, PhysRevB.106.205109, Graham2024, PhysRevLett.131.016901} and a large anomalous Hall conductivity \cite{doi:10.1126/sciadv.abb6003, PhysRevB.104.L041103, Xu_2025}. Recent advances in exfoliation techniques have isolated AV$_3$Sb$_5$ down to a few layers \cite{pub.1165232679}, placing the fabrication of twisted bilayer devices within experimental reach. Theoretical studies also confirm that the requisite LC order is stabilized by realistic interactions in their monolayer \cite{PhysRevB.107.045127, Kim2023, IM202326}, so that twisted vanadium-based kagome metals would provide a promising platform for engineering emergent topology and quantum geometry. 

\par Periodic laser driving using a rectangular waveguide can effectively reduce $t_z$  \cite{PhysRevB.101.241408, PhysRevB.103.014310, PhysRevB.104.195429, PhysRevResearch.2.033494, RODRIGUEZVEGA2021168434}, allowing access to the small-$t_z$ regime without changing the underlying material platform, providing a direct pathway for observing the crossover from non-inherited to monolayer-inherited quantum geometry in the moiré flat bands.

\paragraph{Discussion---}Using a TB framework free from the limitations of low-energy continuum models, we have demonstrated that quantum geometric inheritance cannot be taken for granted in moiré systems. In tb-LCK, the interplay between the LC phase and interlayer tunneling provides an independent handle for suppressing, reorganizing, or entirely erasing inherited quantum geometry. These findings carry fundamental implications for moiré quantum matter. Unlike standard tb-TMDs or TBG, where weak coupling allows valley properties to cleanly dictate moiré topology, tb-LCK realizes a new regime where interlayer hybridization itself acts as the primary topological tuning parameter. Our study establishes a universal mechanism that transcends kagome systems, pointing to entirely new classes of moiré phenomena that lie beyond the reach of the topological inheritance paradigm.

\paragraph*{Acknowledgement---} The work was supported by the National Science Foundation through the Expand-QISE award NSF-OMA-2329067 and benefited from the resources of Northeastern University’s Advanced Scientific Computation Center, the Discovery Cluster, the Massachusetts Technology Collaborative award MTC-22032, and the Quantum Materials and Sensing Institute (QMSI).

\paragraph*{Data availability---}The data that support the findings of this article are not publicly available upon publication because it is not technically feasible and/or the cost of preparing, depositing, and hosting the data would be prohibitive within the terms of this research project. The data are available from the authors upon reasonable request.


\bibliography{apssamp}
\renewcommand{\theequation}{S\arabic{equation}}
\renewcommand{\thefigure}{S\arabic{figure}}
\renewcommand{\thetable}{S\arabic{table}}
\renewcommand{\bibnumfmt}[1]{[S#1]}
\renewcommand{\citenumfont}[1]{S#1}
\newcommand{\bk}{\boldsymbol\kappa}

\newcommand{\beginsupplement}{%
  \setcounter{equation}{0}
  \renewcommand{\theequation}{S\arabic{equation}}%
  \setcounter{table}{0}
  \renewcommand{\thetable}{S\arabic{table}}%
  \setcounter{figure}{0}
  \renewcommand{\thefigure}{S\arabic{figure}}%
  \setcounter{section}{0}
  \renewcommand{\thesection}{S\Roman{section}}%
  \setcounter{subsection}{0}
  \renewcommand{\thesubsection}{S\Roman{section}.\Alph{subsection}}%
}

\clearpage
\pagebreak
\begin{widetext}
\begin{center}
\textbf{\large Supplemental Material: Breakdown of Topological Inheritance and Twist-Induced Quantum Geometry Reconfiguration in Moiré Flat Bands}
\end{center}
\tableofcontents

\section{S1. Reduced Hall conductivity and modified quantum weight}
\par The reduced Hall conductivity $\bar{\sigma}$ is related to the Hall conductivity by $\sigma_{xy}=(e^2/h)\bar{\sigma}$, and is equivalent to the Chern number once it is quantized. It thus captures information about topological phase transitions. The modified quantum weight $\tilde{K}$ generalizes the quantum weight $K$ commonly invoked in the study of optical sum rules \cite{PhysRevX.14.011052, PhysRevLett.133.206602}:
\begin{equation}
K = 2\pi\int d[\mathbf{k}] \text{tr}[g_{\mu\nu}(\mathbf{k})],
\end{equation}
where $g_{\mu\nu}(\mathbf{k})$ with $\mu,\nu=x,y$ is the quantum metric and $\text{tr}[...]$ sums over spatial indices. Using the positive semi-definiteness of the quantum geometric tensor, it can be readily shown that \cite{PhysRevB.90.165139, PhysRevB.95.024515, Peotta2015, PhysRevLett.124.167002, PhysRevLett.128.087002}:
\begin{equation}\label{eq:S2}
    \text{tr}(g_{\mu\nu}(\mathbf{k})) \geq 2\sqrt{\text{det}(g_{\mu\nu}(\mathbf{k}))} \geq |\Omega_{xy}(\mathbf{k})|.
\end{equation}
Therefore, the modified quantum weight $\tilde{K}$ provides a lower bound for the quantum weight $K$.

\par To prove Eq.~\eqref{eq:S2}, we first introduce the quantum geometric tensor with matrix elements\cite{Yu2025}:
\begin{equation}\label{eq:S3}
Q_{\mu\nu}(\mathbf{k}) = \text{Tr}[(\partial_\mu P(\mathbf{k}))\bar{Q}(\mathbf{k})(\partial_\nu P(\mathbf{k}))],
\end{equation}
where $\mu,\nu=x,y$ are the spatial indices, $P(\mathbf{k})$ is the projection operator of the manifold of interest, $\bar{Q}(\mathbf{k})\equiv1-P(\mathbf{k})$, $\partial_\mu\equiv\partial/\partial k^\mu$, and $\text{Tr}[...]$ sums over quantum states. The quantum geometric tensor defined in Eq.~\eqref{eq:S3} is positive semi-definite since:
\begingroup\makeatletter\def\f@size{9.5}\check@mathfonts
\def\maketag@@@#1{\hbox{\m@th\large\normalfont#1}}
\begin{equation}
\begin{split}
(v^\mu)^\dagger Q_{\mu\nu}(\mathbf{k})v^\nu & = (v^\mu)^\dagger\text{Tr}[\partial_{\mu}P(\mathbf{k})\bar{Q}(\mathbf{k})\partial_{\nu}P(\mathbf{k})]v^\nu
\\ & = \text{Tr}[ ((\bar{Q}(\mathbf{k})\partial_{\nu}P(\mathbf{k}))v^\nu)^\dagger((\bar{Q}(\mathbf{k})\partial_{\nu}P(\mathbf{k}))v^\nu) ] 
\\ & \geq 0,
\end{split}
\end{equation}
\endgroup
where $v\in\mathbb{C}^2$ is some vector. Therefore, $\text{det}(Q(\mathbf{k}))\geq0$.  Since $Q(\mathbf{k})$ is a $2\times2$ Hermitian matrix such that $\det(Q(\mathbf{k}))=\det(\text{Re}(Q(\mathbf{k})))-\det(\text{Im}(Q(\mathbf{k})))$, it leads to $\det(\text{Re}(Q(\mathbf{k})))\geq\det(\text{Im}(Q(\mathbf{k})))$ due to the positive semi-definiteness of $Q(\mathbf{k})$. Note that, $\det(\text{Re}(Q(\mathbf{k})))=\det(g(\mathbf{k}))$ and $\det(\text{Im}(Q(\mathbf{k})))=\frac{1}{4}\Omega_{xy}^2(\mathbf{k})$, as $Q_{\mu\nu}(\mathbf{k})=g_{\mu\nu}(\mathbf{k})+i\Omega_{\mu\nu}(\mathbf{k})/2$ \cite{Yu2025}. This proves the second inequality in Eq.~\eqref{eq:S2}. The first one in Eq.~\eqref{eq:S2} can be readily proved by using the AM-GM inequality.

\par To encompass the case of conducting bands, we set
\begin{equation}
\Omega_{xy}(\mathbf{k})=\sum_{n\in M}f_n(T,\mu)\Omega_{n,xy}(\mathbf{k}), 
\end{equation}
where $M$ indicates bands of interest, $f_n(T,\mu)$ is the Fermi-Dirac distribution of the nth band at temperature $T$ and chemical potential $\mu$, and $\Omega_{n,xy}(\mathbf{k})$ is the Berry curvature of the nth band. This is used in the calculations shown in Fig.~5 (d) of the main text.

\par As noted in the main text, for $t_z=0.03|t|$, $\bar{\sigma}$ reverses sign at low temperature ($k_BT\approx10^{-7}|t|$ [Fig.~\ref{fig:S12}) includes temperature effects of redistribution of Berry curvature contributions]. The magnitudes of both $\bar{\sigma}$ and $\tilde{K}$, however, remain enhanced compared to the values for $t_z=0.3|t|$ at the same twist angle ($\theta_c\approx3.89^\circ$).

\begin{figure}[t]
  \centering
  \centering
    \includegraphics[width=\linewidth]{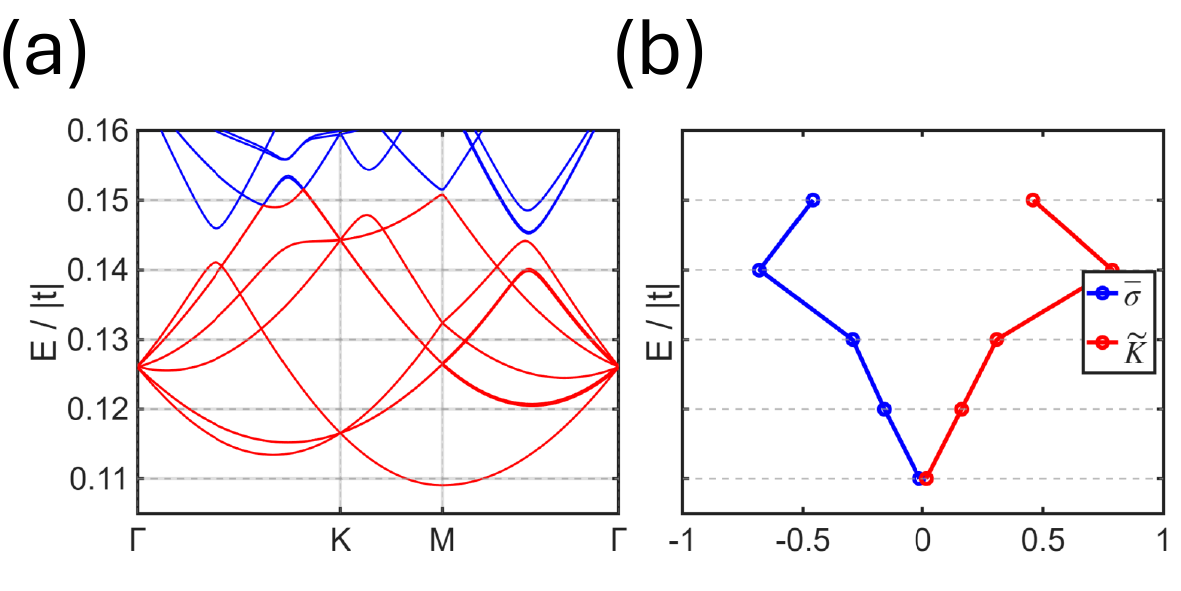}
  \caption{ (a) Band structure of tb-LCK for $\phi_{\text{LC}} = 0.3976\pi$ and $t_z=0.03|t|$ at $\theta_c\approx3.89^\circ$, where the bands of interest are highlighted in red lines and labeled by their lower bound of quantum weight $\tilde{K}$. (b) The $\bar{\sigma}$ and $\tilde{K}$ of tb-LCK with $t_z=0.03|t|$ at $\theta_c\approx3.89^\circ$ computed at selective chemical potentials marked by gray dashed lines in (a) at $k_BT\approx8.6\times10^{-7}|t|$.
  }
  \label{fig:S12}
\end{figure}

\section{S2. High-symmetry bilayer stackings of LCK}
\par Here we discuss the rotational symmetry axes in LCK and the structure of its two high-symmetry stacked bilayers with reference to Figs.~\ref{fig:S1} and \ref{fig:S2}. These bilayers with $C_{2z}$- and $C_{3z}$-symmetries are generated by translating the $C_{6z}$-symmetric center of the first layer to the other $C_{nz}$-symmetric centers of of the second layer, which reduces the overall rotational symmetry of the bilayer to $C_{nz}$. A similar high-symmetry pattern is also found in twisted bilayer kagome lattices \cite{PhysRevB.100.155421, Phys.Rev.B.112.035410, Phys.Rev.B.112.235134}. For illustrative purposes, we provide one example each of bilayer systems in Fig.~\ref{fig:S2}. Note that the $C_{6z}$-symmetric bilayer is simply the AA-stacked bilayer. 

\begin{figure}[h]
  \centering
  \centering
    \includegraphics[width=\linewidth]{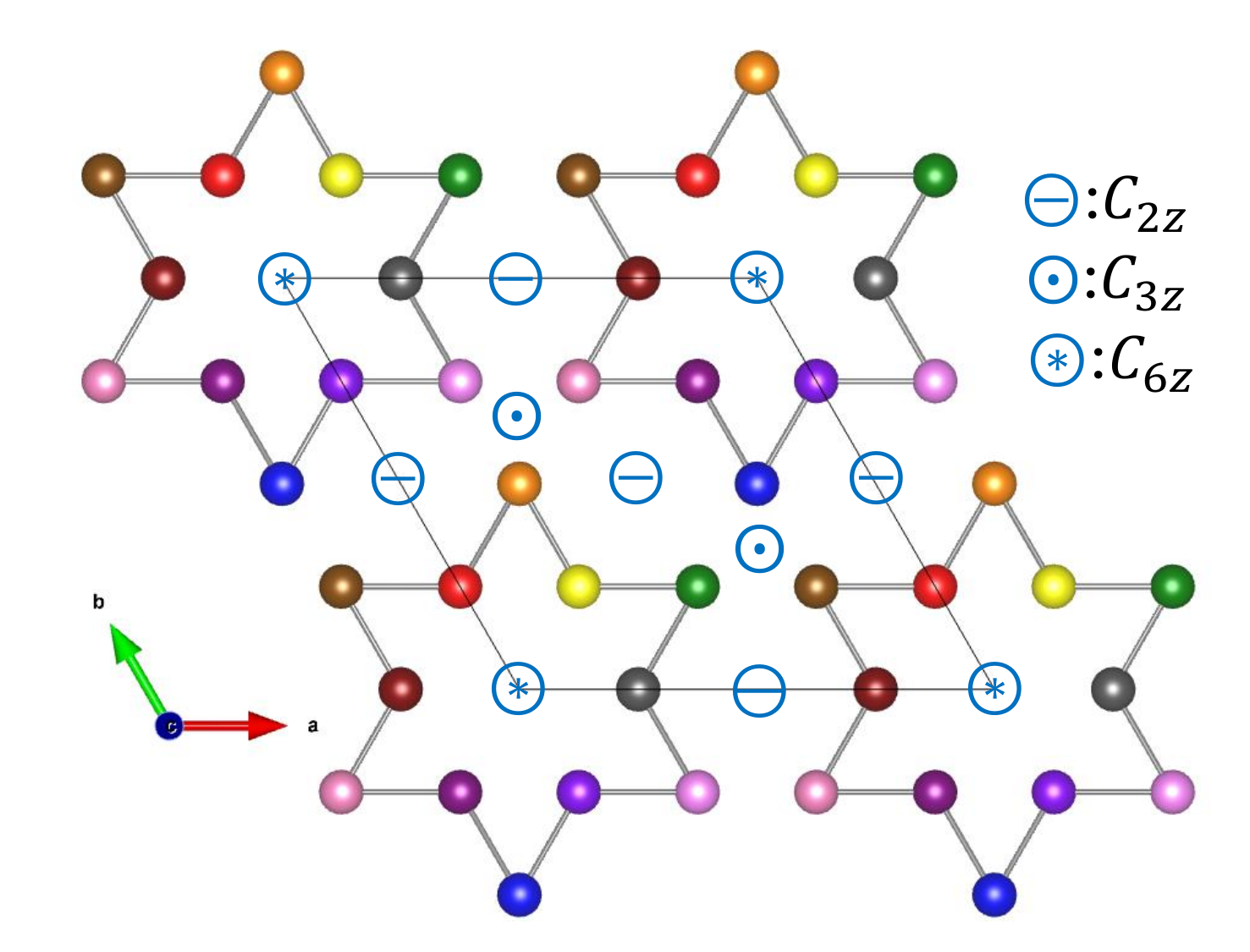}
  \caption{ Lattice structure of LCK. The $C_{nz}$-symmetric centers are labeled.
  }
  \label{fig:S1}
\end{figure}

\begin{figure}[h]
  \centering
  \centering
    \includegraphics[width=\linewidth]{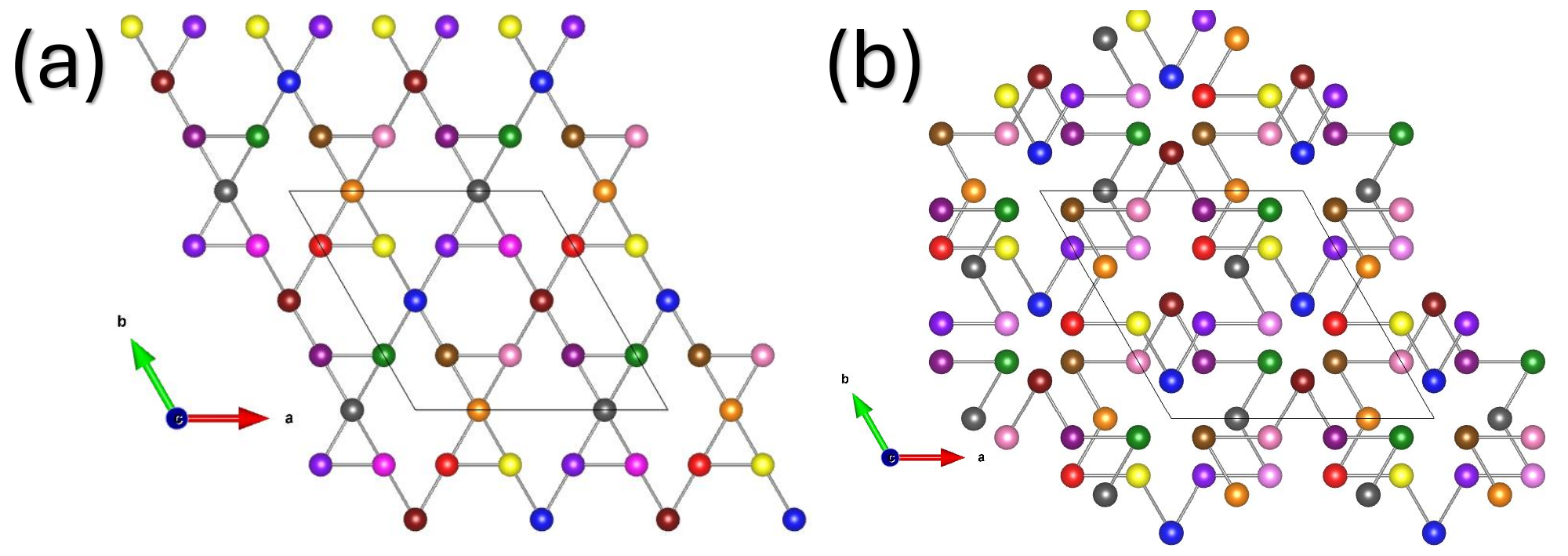}
  \caption{ (a) $C_{2z}$- and (b) $C_{3z}$-symmetric stacked bilayers.
  }
  \label{fig:S2}
\end{figure}

\section{S3. Electronic structure of LCK for various values of $\phi_{\text{LC}}$}
\par We present the evolution of band structure and Berry curvature distribution of band B6 in LCK for $\phi_{\text{LC}}$ values not shown in the main text (Fig.~\ref{fig:S3}). Note that TRS is preserved for $\phi_{\text{LC}} = 0$ with real $\pi$. These results make it clear that the topology of band B6 originates from multiple band inversions at the $\Gamma$ point upon the addition of the LC order (Fig.~\ref{fig:S4}).

\begin{figure}[t]
  \centering
  \centering
    \includegraphics[width=\linewidth]{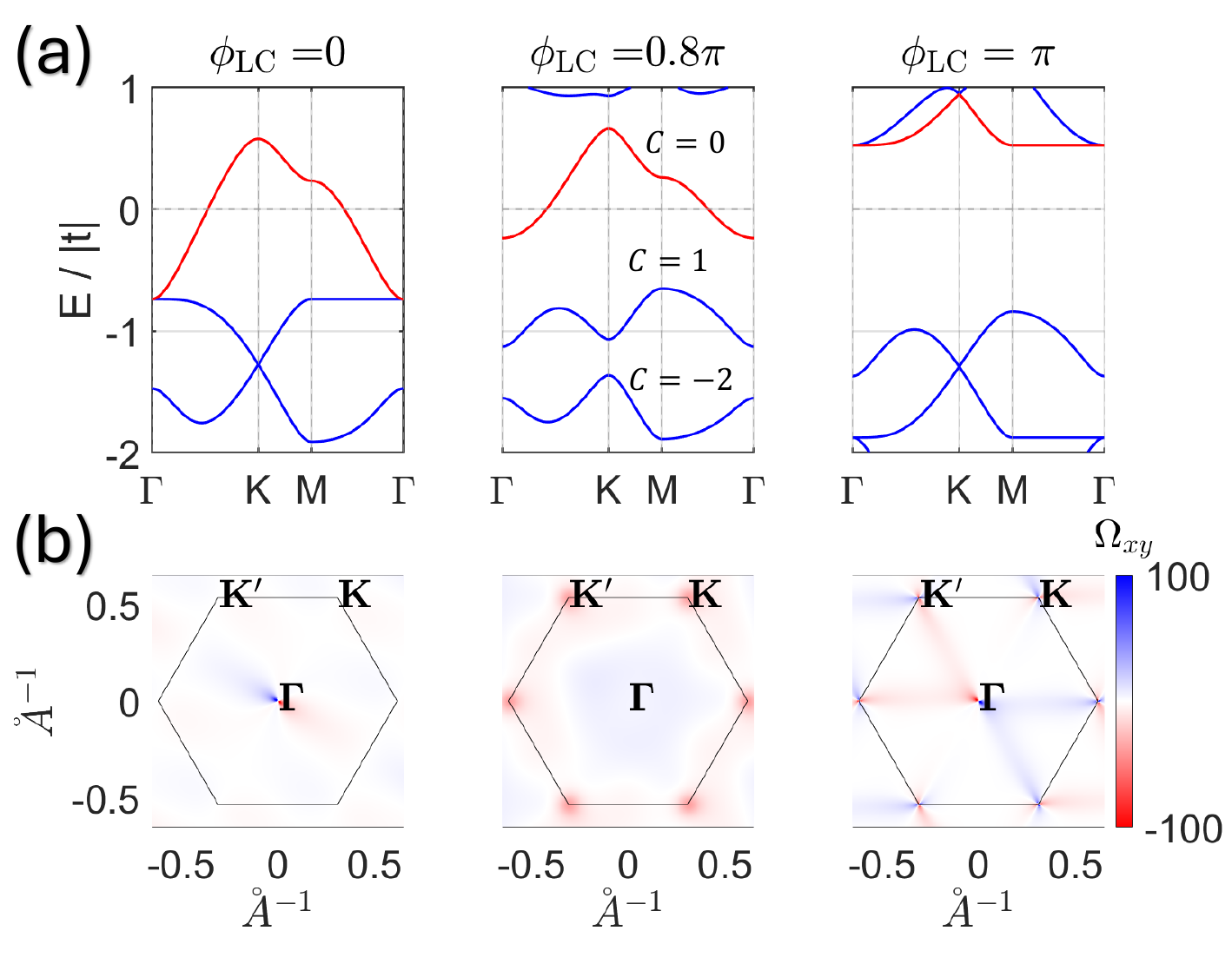}
  \caption{ (a) Band structures of LCK at $\phi_{\text{LC}} = 0$, $0.8\pi$, and $\pi$, with the 6th lowest band (B6) highlighted in red line and nearby bands labeled by their Chern numbers. (b) Berry curvatures of band B6 for various $\phi_{\text{LC}}$ values.
  }
  \label{fig:S3}
\end{figure}

\begin{figure}[t]
  \centering
  \centering
    \includegraphics[width=\linewidth]{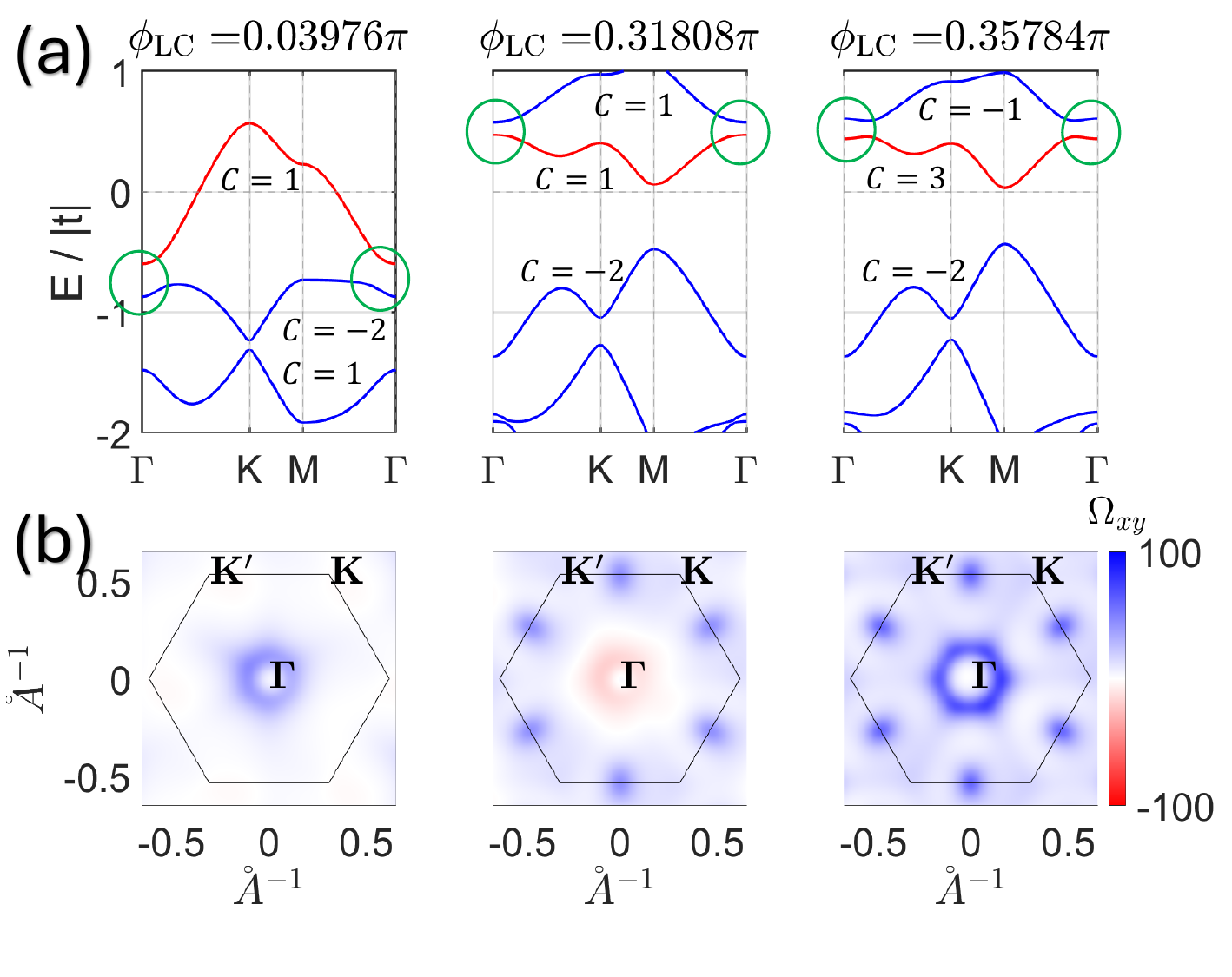}
  \caption{ (a) Band structures of LCK at $\phi_{\text{LC}} = 0.1\phi_0$, $0.8\phi_0$, and $0.9\phi_0$, for $\phi_0=0.3976\pi$. The 6th lowest band (B6) is marked in red line and nearby bands are labeled by their Chern numbers. Green frames highlight the band inversion points. (b) Berry curvatures of band B6 for various $\phi_{\text{LC}}$ values.
  }
  \label{fig:S4}
\end{figure}

\section{S4. Electronic structure of tb-LCK for various $\phi_{\text{LC}}$} values
\par We present the band structure of tb-LCK for $\phi_{\text{LC}} values of 0$, $0.8\pi$, and $\pi$ (Fig.~\ref{fig:S5}); these results are not shown in the main text. $\nu=\nu_{\nu_0}\pm n$ indicates the filling near $\nu_0$ with additional $\pm n$ electrons. Although the flat bands are generated here from the band B6 of the monolayer, these flat bands are featureless from the viewpoint of quantum geometry, like the monolayer they originate from (Fig.~\ref{fig:S3}). Berry curvature distributions and the Wilson-loop spectra of topologically non-trivial bands in tb-LCK for $\phi_{\text{LC}}=0.2\pi$ at $\theta_c\approx3.89^\circ$ are shown in Fig.~\ref{fig:S6}.

\begin{figure}[t]
  \centering
  \centering
    \includegraphics[width=\linewidth]{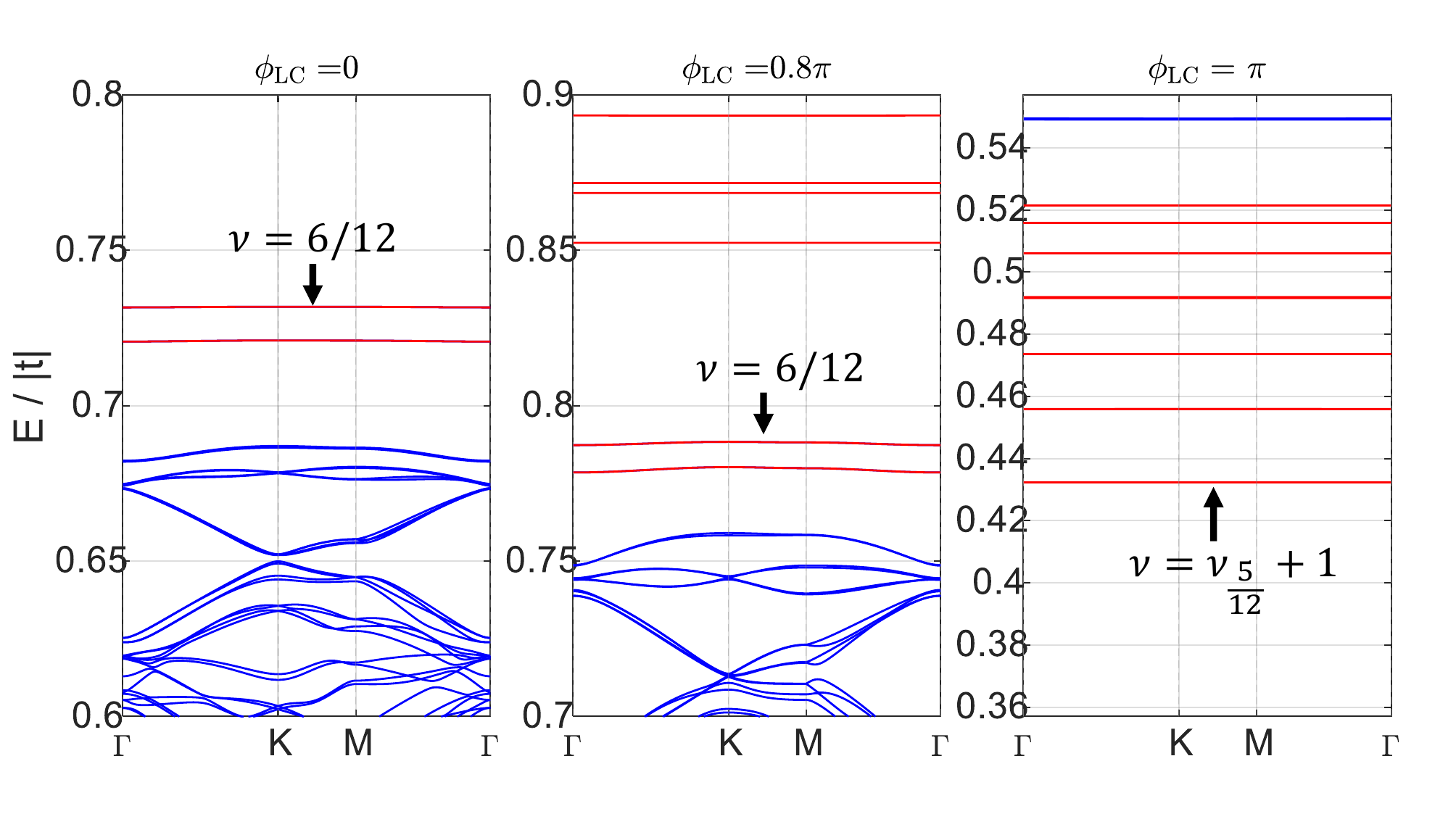}
  \caption{ Band structures of LCK for $\phi_{\text{LC}} values of 0\pi$, $0.8\pi$, and $\pi$ at $\theta_c\approx3.89^\circ$, with the bands of interest highlighted in red lines. The filling factor $\nu$ is indicated.
  }
  \label{fig:S5}
\end{figure}

\begin{figure}[t]
  \centering
  \centering
    \includegraphics[width=\linewidth]{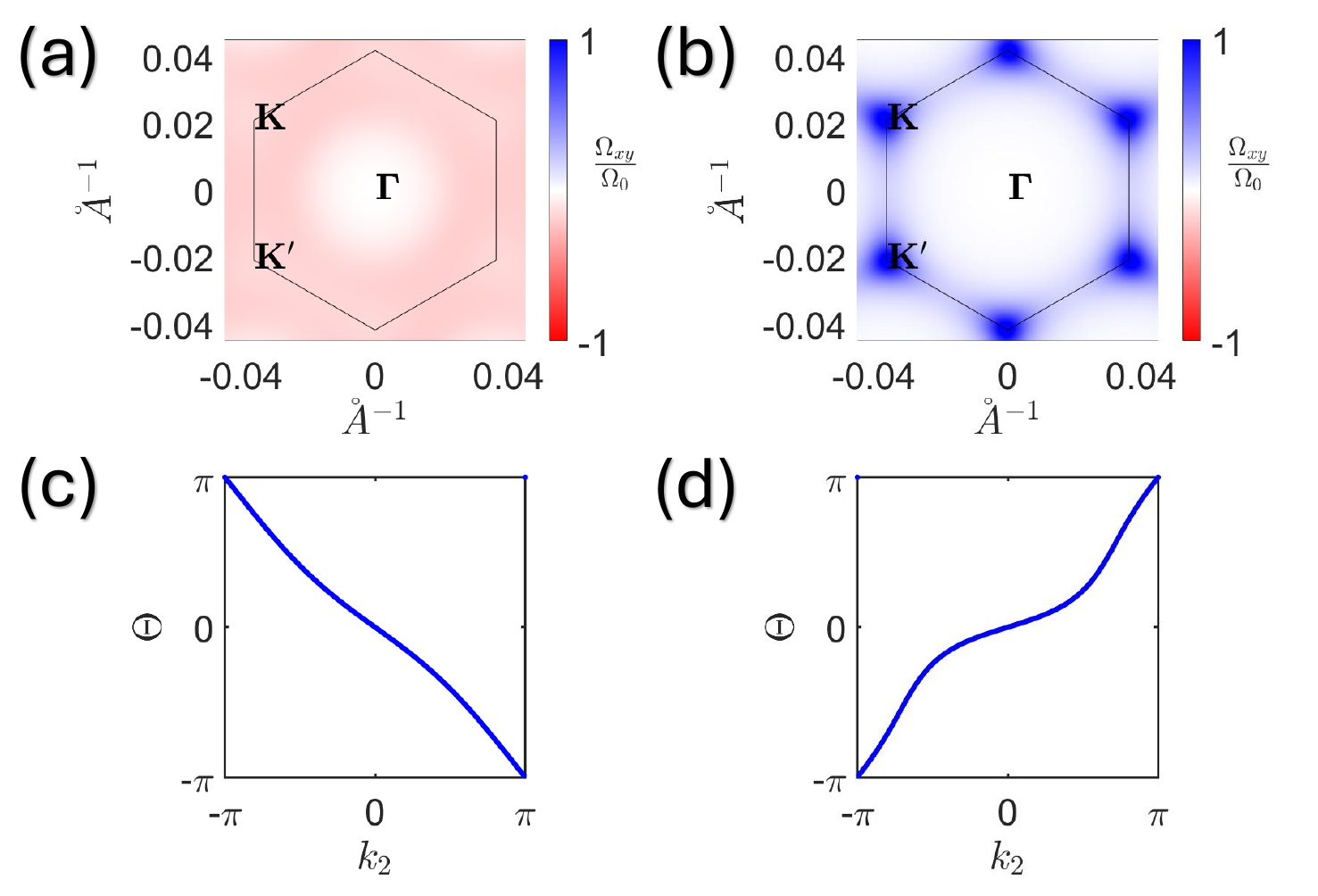}
  \caption{ Berry curvature distribution of selected bands in tb-LCK at $\theta_c\approx3.89^\circ$ scaled by $\Omega_0=10^4$. (a, b) Bands with $\nu = \nu_{5/12} + 4$ and $\nu = \nu_{5/12} + 6$ at $\phi_{\text{LC}} = 0.2\pi$, and (c,d) the corresponding Wilson-loop spectra.
  }
  \label{fig:S6}
\end{figure}

\section{S5. Electronic structure of tb-LCK for $\phi_{\text{LC}}\approx0.3976\pi$ at various twist angles}
We present tb-LCK results for  $\phi_{\text{LC}}\approx0.3976\pi$ at various twist angles, including the band structures, Wilson-loop spectra, and Berry curvature distributions. Fig.~\ref{fig:S7} shows the 12 low-energy bands that originate from the lower band edge of band B6 at commensurate twist angles ranging from $\theta_c=13.17^\circ$ to $\theta_c=3.48^\circ$. As the twist angle decreases, the bands are seen to progressively cluster. At $\theta_c = 6.01^\circ$ and $5.09^\circ$, they split into two groups: a cluster of two bands and another cluster of ten bands. The latter cluster further separates into four- and six-band subclusters near $\theta_c \approx 4.41^\circ$, and the four-band group isolates a single band at $\theta_c \leq 3.89^\circ$. Since the bands within each cluster remain entangled, we evaluate their modified quantum weight $\tilde{K}$ collectively.

\par Fig.~\ref{fig:S8} presents the Berry curvature distribution of the twelve low-energy bands that originate from the lower band edge of band B6 at $\theta_c\approx13.17^\circ$, along with the corresponding Wilson-loop spectrum.

\begin{figure}[h]
  \centering
  \centering
    \includegraphics[width=\linewidth]{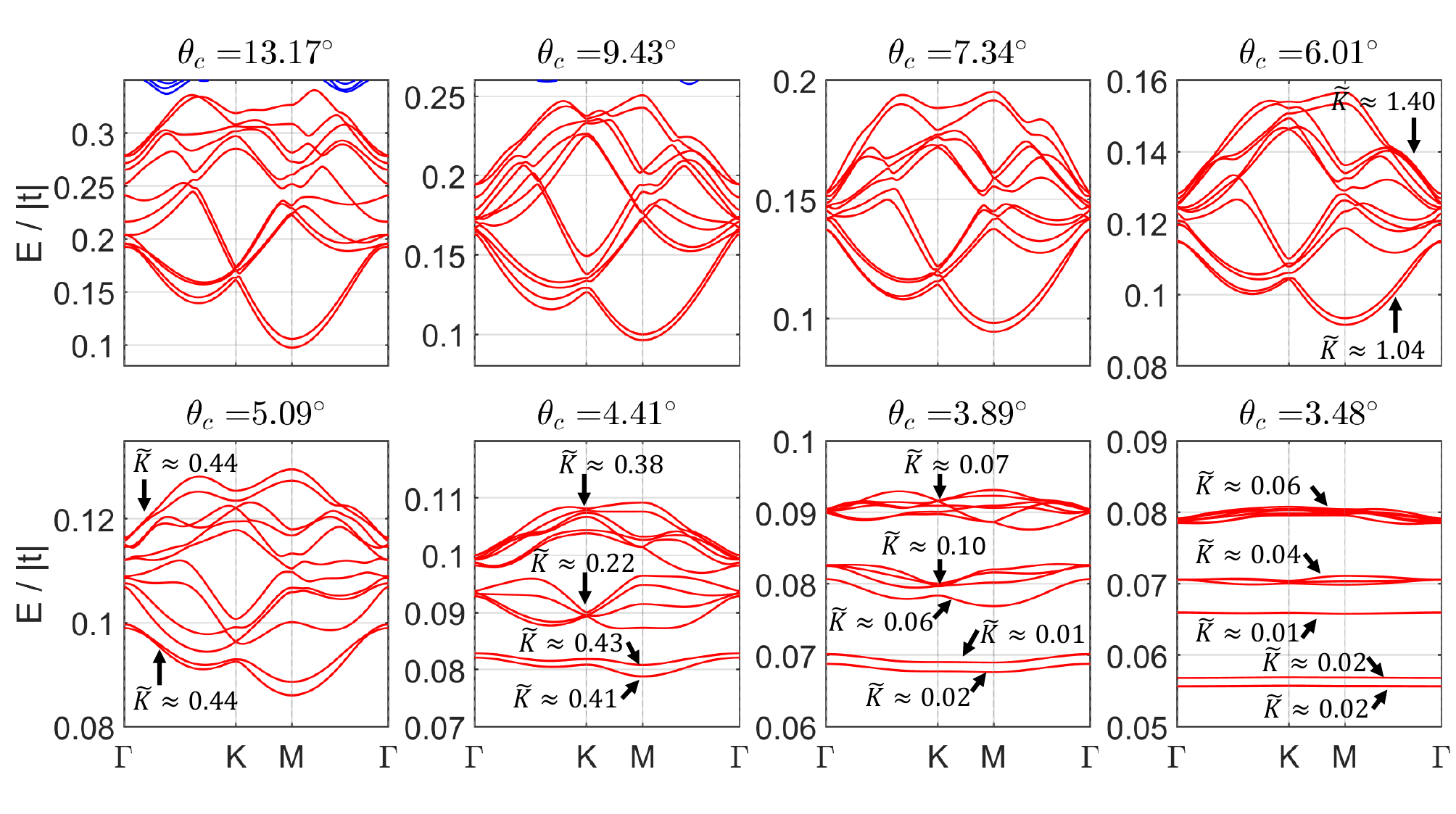}
  \caption{ Band structures of LCK for $\phi_{\text{LC}} values of 0.3976\pi$ and $0.8\pi$ at various commensurate twist angles, with the bands of interest highlighted in red lines. Band clusters are also labeled by their modified quantum weights $\tilde{K}$.
  }
  \label{fig:S7}
\end{figure}

\begin{figure}[h]
  \centering
  \centering
    \includegraphics[width=\linewidth]{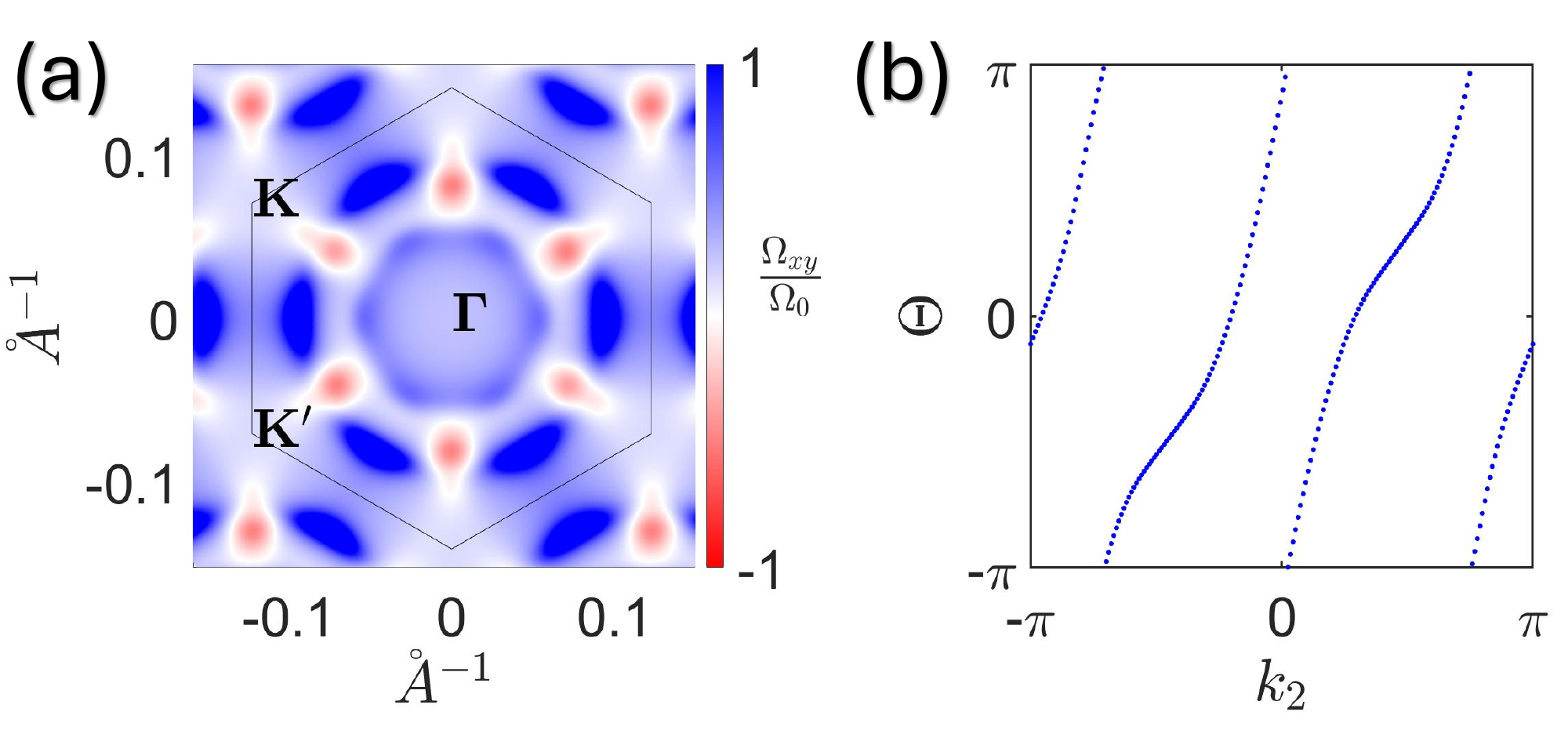}
  \caption{ (a) Berry curvature distribution of the 12 low-energy bands in tb-LCK for $\phi_{\text{LC}} = 0.3976\pi$ at $\theta_c\approx13.17^\circ$ scaled by $\Omega_0=10^3$, and (b) the corresponding Wilson-loop spectrum.
  }
  \label{fig:S8}
\end{figure}

\section{S6. Electronic structure of LCK and tb-LCK for $\phi_{\text{LC}}\approx0.6\pi$}
Band structures of LCK and tb-LCK at $\theta\approx3.89^\circ$ are presented in Figs.~\ref{fig:S9} (a) and (c); band B6 and the associated interesting moiré flat bands are colored by red lines and labeled by their Chern numbers. Note that although band B6 is topologically trivial, it possesses a concentrated Berry curvature at the $K,K'$ valleys. Due to the broken TRS, the Berry curvature at both valleys has the same strength. Together with the strong hybridization with the 7th lowest bands from the interlayer tunneling, this suggests a potential origin of the high Chern numbers of flat Chern bands in tb-LCK. Figs.~\ref{fig:S9} (b) and \ref{fig:S10} present the Berry curvature of band B6 in LCK and the bands of interest in tb-LCK, along with their Wilson-loop spectra.

\begin{figure}[ht]
  \centering
    \includegraphics[width=\linewidth]{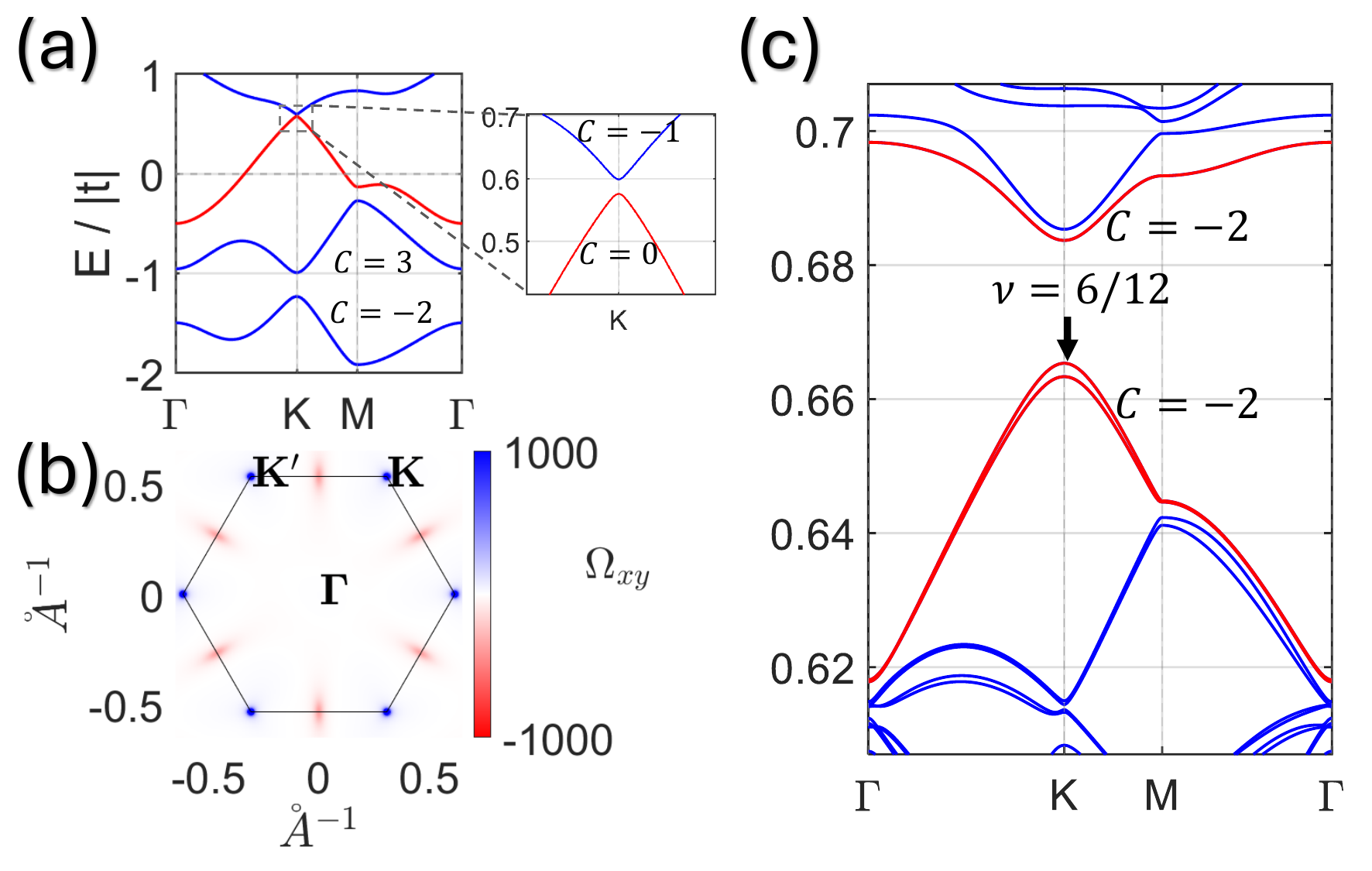}
  \caption{ (a) Band structure of LCK for $\phi_{\text{LC}} = 0.6\pi$. Band B6 is given in red line and labeled by its Chern number. Inset shows a zoom-in view of the gapped Dirac cone. (b) The corresponding Berry curvature distribution of band B6. (c) Band structure of tb-LCK for $\phi_{\text{LC}} = 0.6\pi$ at $\theta_c\approx3.89^\circ$; red lines identify bands of interest labeled by their Chern number.
  }
  \label{fig:S9}
\end{figure}
\begin{figure}[h]
  \centering
    \includegraphics[width=\linewidth]{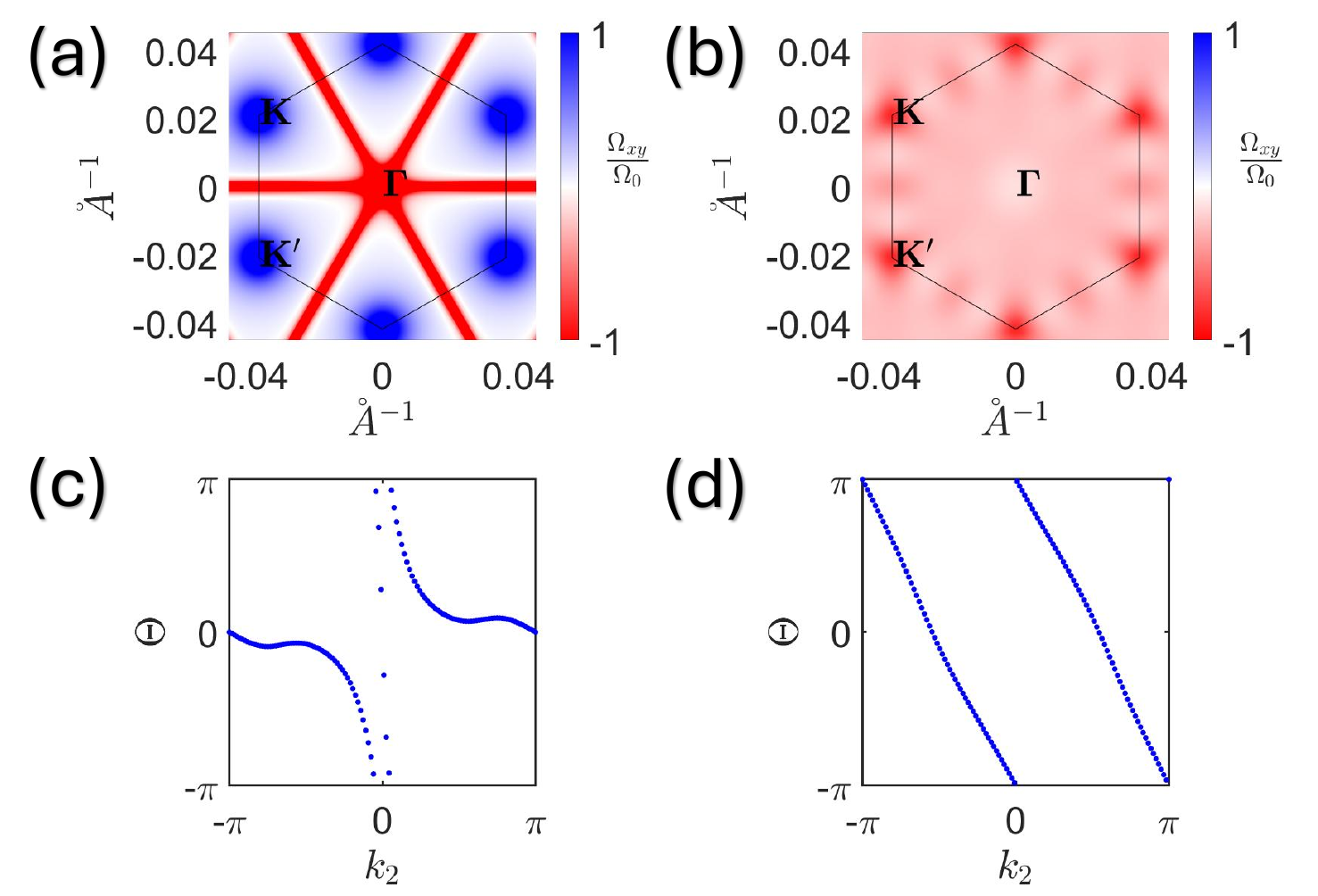}
  \caption{ (a,b) Berry curvature distribution of selected bands in tb-LCK for $\phi_{\text{LC}} = 0.6\pi$ at $\theta_c\approx3.89^\circ$ scaled by $\Omega_0=10^4$, corresponding to bands with ($\nu = \nu_{6/12} - 1$,$\nu = \nu_{6/12}$) and $\nu = \nu_{6/12} + 1$, following the color scheme of Fig.~\ref{fig:S9}. (c,d) The corresponding Wilson-loop spectra.
  }
  \label{fig:S10}
\end{figure}

\section{S7. Electronic structure of LCK and tb-LCK for $\phi_{\text{LC}} \approx0.2\pi$ and $t_z=0.03|t|$}
\par Fig.~\ref{fig:S11} presents the Berry curvature distributions and the corresponding Wilson-loop spectra of low-energy bands highlighted in Fig.~4 of the main text.
\begin{figure}[h]
  \centering
  \centering
    \includegraphics[width=\linewidth]{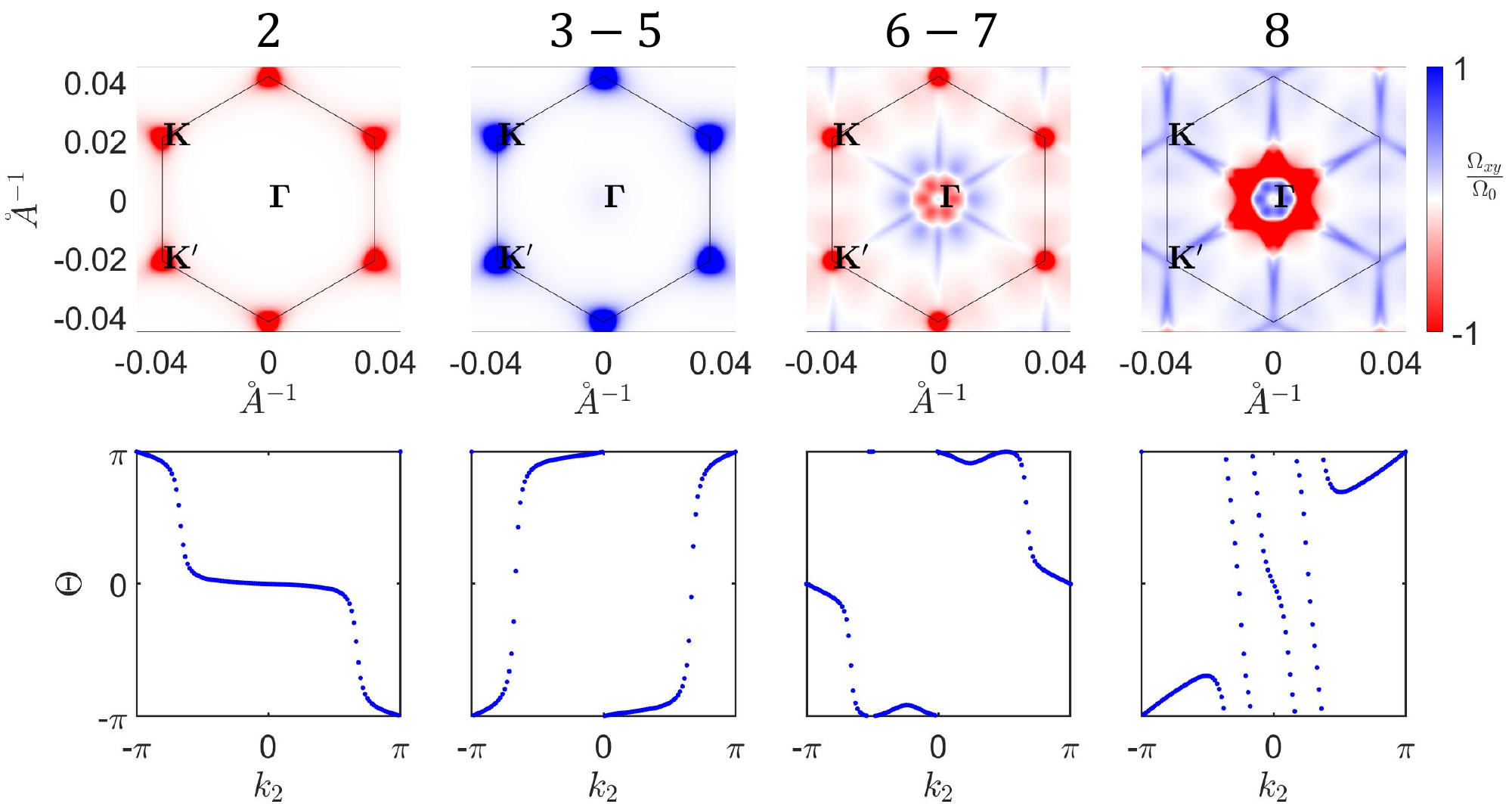}
  \caption{ Berry curvature distribution of low-energy bands in tb-LCK for $\phi_{\text{LC}}\approx0.2\pi$ at $\theta_c\approx3.89^\circ$ scaled by $\Omega_0=10^4$. Band indices from Fig.~4 of the main text are given at the top of various columns. 
  }
  \label{fig:S11}
\end{figure}

\end{widetext}
\end{document}